\documentclass{emulateapj}

\usepackage{graphicx}  
\usepackage{dcolumn}   
\usepackage{bm}        
\usepackage{amsfonts,amsmath,amssymb,mathrsfs}
\usepackage{color}

\usepackage{natbib,times}
\newcommand{\eq}{\begin{equation}}
\newcommand{\eeq}{\end{equation}}
\newcommand{\be}{\begin{equation}}
\newcommand{\ee}{\end{equation}}
\newcommand{\bea}{\begin{eqnarray}}

\newcommand{\eea}{\end{eqnarray}}
\newcommand{\vta}[1]{\vert \boldsymbol{a}_{#1}        \vert}

\newcommand{\vtl}   {\vert \boldsymbol{       {\ell}} \vert}

\newcommand{\eg}{\textit{e.g.,}~}
\newcommand{\ie}{\textit{i.e.,}~}
\newcommand{\cf}{\textit{c.f.}~}

\shorttitle{Predicting the direction of the final spin}
\shortauthors{Barausse \& Rezzolla}

\begin{document}

\title{Predicting the direction of the final spin from the coalescence of two black holes}

\author{Enrico Barausse}
\affil{Center for Fundamental Physics, University of Maryland, College Park, MD 20742-4111, USA}

\and

\author{Luciano Rezzolla}
\affil{Max-Planck-Institut f\"ur Gravitationsphysik, Albert-Einstein-Institut, Potsdam-Golm, Germany}
\affil{Department of Physics and Astronomy, Louisiana State University, Baton Rouge, LA, USA}

\begin{abstract}
Knowledge of the spin of the black hole resulting from
the merger of a generic black-hole binary is of great
importance for studying the cosmological evolution of 
supermassive black holes. Several attempts have been made
to model the spin via simple expressions
exploiting the results of numerical-relativity simulations. While
these expressions are in reasonable agreement with the
simulations, 
they neglect the precession of the binary's orbital plane, and cannot therefore be applied directly 
-- \textit{i.e.,} without evolving the system to small separations using post-Newtonian theory -- 
to binaries with separations larger than a few hundred gravitational radii. While not a problem in principle,
this may be impractical if the formulas are employed in cosmological merger-trees or
N-body simulations, which provide the spins and angular momentum of the two
black holes when their separation is of hundreds or thousands of gravitational radii. The formula that we propose
is instead built on improved assumptions and gives,
for any separation, a very accurate prediction both for the norm of
the final spin and for its direction. By comparing with the
numerical data, we also show that the final-spin direction
is very accurately aligned with the binary's total angular momentum at large
separation. Hence, observations of the final-spin direction (\eg
via a jet) can provide information on the binary's orbital plane
at large separations and could be relevant, for instance, for studying X-shaped
radio sources.
\end{abstract}

\keywords{black-hole physics --- relativity --- gravitational waves ---
galaxies: nuclei}
\maketitle

\section*{Introduction}

While analytic solutions for isolated black holes (BHs) have a long
history, the dynamics of BH binaries has
been solved only recently and through computationally-expensive
numerical-relativity (NR) calculations [see~\citet{NR} for a review]. Despite the mathematical complexity of the
problem, many results of the NR
simulations can be reproduced accurately using semi-analytical
prescriptions~\citep{eob1,eob2} based on post-Newtonian (PN) and BH
perturbation theory. It is therefore not entirely surprising that the
dimensionless spin of the remnant from a BH-binary merger,
$\boldsymbol{a}_{\rm fin}=\boldsymbol{S}_{\rm fin}/M_{\rm fin}^2$, can
be described, with increasing accuracy, via simple prescriptions based
on point particles~\citep{scott,bkl,kesden}, on fits to the NR
data~\citep{fit1,fit2,faunew,boyle1,boyle2}, or on a combination of
the two approaches~\citep{old} [see~\citet{lr09} for a review]. These formulas are useful because they
provide information over the entire 7-dimensional parameter space 
for BH binaries in quasi-circular orbits, namely: the 
mass ratio $q\equiv M_2/M_1$ and the six components of the initial
dimensionless spin vectors
$\boldsymbol{a}_{1,2}=\boldsymbol{S}_{1,2}/M_{1,2}^2$. Such parameter
space could in principle be investigated entirely via NR calculations;
in practice, however, the simulations are still very expensive and
restricted to $q=0.1$--$1$. Also, these formulas have many 
applications: in astrophysics, where they could provide
information on massive star binaries~\citep{fabian}; in cosmology, where supermassive BHs (SMBHs) are
believed to assemble through accretion and
mergers~\citep{cosmo_spin_ev}; in gravitational-wave astronomy, where
\textit{a priori} knowledge of the final spin can 
help the detection~\citep{ringdown}.

While the different expressions for the spin norm,
$|\boldsymbol{a}_{\rm fin}|$, are in good agreement among themselves
and with the numerical data, the predictions for the final-spin direction, 
$\hat{\boldsymbol{a}}_{\rm fin}\equiv {\boldsymbol{a}}_{\rm fin}/|{\boldsymbol{a}}_{\rm fin}|$, do not agree well with one another. 
Moreover, all expressions are built from and model the typical ``NR
binaries'' and hence the dynamics of the last few orbits before the
merger, and do not account systematically for the precession
of the orbital angular momentum $\boldsymbol{L}$, thus becoming imprecise when the binary is widely separated~\citep{faunew}.
Of course, it is possible to use the PN equations and evolve a
widely-separated binary to small separations, read-off the relevant
information and apply the presently-available formulas [this was suggested
by~\cite{faunew}, who correctly remark that their expression is valid only at small
separations].
While not a problem in principle, this procedure can be impractical in 
applications, such as
cosmological merger-trees or N-body simulations, that provide the spins 
of the two BHs at separations of hundreds or thousands of gravitational
radii. 

We follow here a different approach, and using assumptions slightly different from those made
in~\citet{old}, we present a new expression for
$\boldsymbol{a}_{\rm fin}$ which is applicable to binaries with arbitrary separations and which also provides better results
for the final-spin direction at small separations. 

\section*{Derivation of the formula}

We recall that
when the BHs have spins that are aligned with $\boldsymbol{L}$, the NR results are accurately described by~\citep{old} 
\begin{equation}
\label{eqspin_uneqmass}
a_{\rm fin}=\tilde{a}
+\tilde{a} \nu (s_{4}\tilde{a}+s_{5}\nu + t_0)
+ \nu(2\sqrt{3}+t_2\nu+t_{3}\nu^2)\,,
\end{equation}
where $\nu \equiv M_1M_2/(M_1+M_2)^2$ is the symmetric mass ratio and
$\tilde{a}\equiv(a_1 + a_2 q^2)/(1+q^2)$. The five coefficients $t_0$,
$t_2$, $t_3$, $s_4$ and $s_5$ in~\eqref{eqspin_uneqmass} can be
determined straightforwardly by fitting the results of the NR
calculations. However, an additional condition can be employed by using
the results obtained by~\citet{caltech_cornell} for
equal-mass non-spinning BHs and thus enforce that for
$a_1=a_2=0,\,\nu=1/4$ and to the claimed precision
\begin{equation}
\label{eq:constraint}
a_{\rm fin}= \frac{\sqrt{3}}{2} + \frac{t_2}{16} + \frac{t_3}{64} 
= 0.68646 \pm 0.00004\,.
\end{equation}
This leaves only \textit{four} unconstrained coefficients, so that by
using the NR results for the 72 \textit{aligned} binaries published so 
far~\citep{fit1, fit2, buonanno, berti1, berti2, bam,
  caltech_cornell} we obtain 
\begin{align}
\label{eq:coeff}
&s_4 = -0.1229\pm0.0075\,,\quad s_5 = 0.4537\pm0.1463\,,\nonumber\\ 
&t_0=-2.8904\pm0.0359\,,\quad t_3 = 2.5763\pm0.4833\,,
\end{align}
with an agreement $\vert a^{\rm NR}_{\rm fin}-a^{\rm fit}_{\rm
  fin}\vert \leq0.0085$ with the data\footnote{Recently, a 
new formula for the final spin has been proposed by~\citet{RITformula}. 
While that formula cannot be easily applied to generic configurations 
(because it depends on the final-plunge direction, which can only be 
determined via simulations), it reduces to a fit for \textit{aligned} 
binaries. For the binaries considered above, it gives a much larger maximum error $\vert a^{\rm NR}_{\rm fin}-a^{\rm fit}_{\rm
fin}\vert \approx 0.05$.}; using the constraint (2) we then
also obtain $t_2=-3.5171 \pm 0.1208$.
Because of 
the larger data set
used, the values~\eqref{eq:coeff} are slightly different from those
in~\citet{old}.

Because~\eqref{eqspin_uneqmass} provides information over only 3 of
the 7 dimensions of the parameter space, we will next show how 
to cover the remaining 4 dimensions and thus derive an expression for
$\boldsymbol{a}_{\rm fin}$ for \textit{generic} BH binaries in
quasi-circular orbits. Following the spirit of~\citet{old}, we make
the following assumptions:

\smallskip
\noindent\textit{(i) The mass $M_{\rm rad}$ radiated to gravitational waves can be neglected} 
\ie $M_{\rm fin} = M \equiv M_1 + M_2$. 
The radiated mass could be accounted for by using the NR data
for $M_{\rm fin}$~\citep{faunew} or extrapolating the
test-particle behavior~\citep{kesden}. The reason why assumption \textit{(i)} is
reasonable is that $M_{\rm rad}$ is largest for aligned
binaries, but these are also the ones employed to fit the free
coefficients~\eqref{eq:coeff}. Therefore, $M_{\rm rad}$ is
approximately accounted for by the values of the coefficients.

\smallskip
\noindent\textit{(ii) The  norms $|\boldsymbol{S}_1|$,
  $|\boldsymbol{S}_2|$, $|\boldsymbol{\tilde{\ell}}|$ do not depend on
  the binary's separation $r$}, with $\boldsymbol{\tilde{\ell}}$ being defined as 
\begin{equation}
\label{ell_def}
\boldsymbol{\tilde{\ell}}(r)\equiv \boldsymbol{S}_{\rm
  fin}-[\boldsymbol{S}_1(r)+\boldsymbol{S}_2(r)]=
\boldsymbol{L}(r)-\boldsymbol{J}_{\rm rad}(r)\,, 
\end{equation} 
where $\boldsymbol{S}_1(r)$, $\boldsymbol{S}_2(r)$ and
$\boldsymbol{L}(r)$ are the spins and the orbital angular momentum
at separation $r$ and $\boldsymbol{J}_{\rm
  rad}(r)$ is the angular momentum radiated from $r$ to the
merger. Clearly, $\boldsymbol{S}_1$, $\boldsymbol{S}_2$ and $\boldsymbol{\tilde{\ell}}$ can still depend on $r$ through their directions. 
While the constancy of $|\boldsymbol{S}_1|$ and $|\boldsymbol{S}_2|$ is a very good
assumption for BHs, which do not have an internal structure,
the constancy of
 $|\boldsymbol{\tilde{\ell}}|$ is more heuristic and based on the
idea that the merger takes place at an ``effective''
innermost stable circular orbit (ISCO), so that $|\boldsymbol{\tilde{\ell}}|$ can be interpreted
as the residual orbital angular momentum contributing
to $\boldsymbol{S}_{\rm fin}$.

\smallskip 
\noindent\textit{(iii) The final spin $\boldsymbol{S}_{\rm fin}$ is
  parallel to the initial total angular momentum
  $\boldsymbol{J}(r_{\rm in})\equiv\boldsymbol{S}_1(r_{\rm in})+
  \boldsymbol{S}_2(r_{\rm in})+\boldsymbol{L}(r_{\rm in})$.} This amounts to assuming that $\boldsymbol{J}_{\rm
  rad}(r_{\rm in}) \parallel \boldsymbol{J}(r_{\rm in})$. It replaces
the assumption, made in~\citet{old}, that $\boldsymbol{J}_{\rm
  rad}(r_{\rm in}) \parallel \boldsymbol{L}(r_{\rm in})$, which is
only valid for a smaller set of configurations. We note that this
assumption is motivated by PN theory: Within the adiabatic approximation,
the secular angular-momentum losses via gravitational radiation are along
$\boldsymbol{J}$~\citep{apostolatos}. This is because as $\boldsymbol{L}$ rotates around
$\boldsymbol{J}$, the emission orthogonal to $\boldsymbol{J}$ averages
out.

\smallskip
\noindent\textit{(iv) The angle between $\boldsymbol{L}$ and
  $\boldsymbol{S}\equiv\boldsymbol{S}_1+ \boldsymbol{S}_2$ and the
  angle between $\boldsymbol{S}_1$ and $\boldsymbol{S}_1$ are constant
  during the inspiral, although $\boldsymbol{L}$ and $\boldsymbol{S}$
  precess around $\boldsymbol{J}$.}

\smallskip
At 2.5 PN order, $\textit{(iii)}$ and
$\textit{(iv)}$ are approximately valid for any mass ratio if only one
of the BHs is spinning, and for $M_1 = M_2$ if one neglects spin-spin couplings.
In both cases, in fact, $\boldsymbol{S}$ and $\boldsymbol{L}$
essentially precess around the direction $\boldsymbol{\hat{J}}$, which remains nearly constant~\citep{apostolatos}, 
and the angle between the two
spins remains constant as well. The only case in which 
$\textit{(iii)}$ and $\textit{(iv)}$ are not even approximately valid is
for binaries which, at some point in the evolution, have
$\boldsymbol{L}(r)\approx-\boldsymbol{S}(r)$. These orbits
undergo the so-called ``transitional precession''~\citep{apostolatos},
as a result of which $\boldsymbol{\hat{J}}$ changes significantly. 
Because transitional precession happens only if $\boldsymbol{L}$ and $\boldsymbol{S}$ are initially \textit{almost}
anti-aligned with $|\boldsymbol{L}|>|\boldsymbol{S}|$, it affects only
a very small region of the parameter space, which is, moreover, poorly
populated if the SMBHs are in a gas-rich environment~\citep{Bogdanovic:2007hp}.  

\begin{figure}
\centerline{
\resizebox{8.cm}{!}{\includegraphics[angle=0]{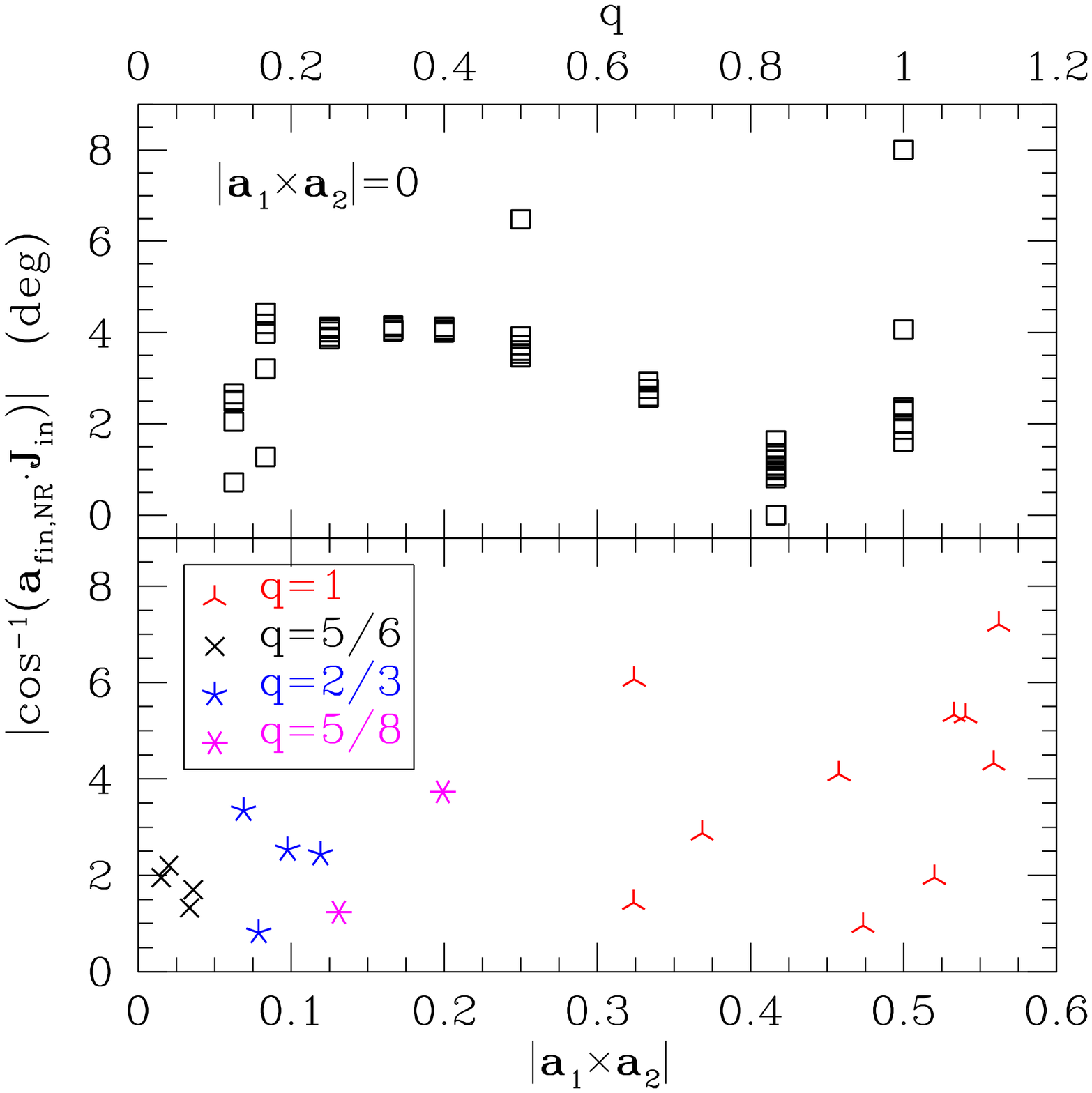}}}
\caption{\small Accuracy of assumption \textit{(iii)} for all the generic binaries published so far, as a
  function of the spin-spin
  coupling ``magnitude'' $|\boldsymbol{a}_1 \times \boldsymbol{a}_2|$ for the binaries with $|\boldsymbol{a}_1 \times \boldsymbol{a}_2|\neq0$
  (lower panel), and of the mass ratio $q$ for the binaries with
  $|\boldsymbol{a}_1 \times \boldsymbol{a}_2|=0$ (upper panel).}
\label{fig:test_assumption3}
\end{figure}

Because assumption \textit{(iii)} plays a major role in our
analysis, in
Fig.~\ref{fig:test_assumption3} we provide evidence of its
(approximate) validity by using
all the \textit{generic} BH-binary simulations (\ie
with spins \textit{not} parallel to $\boldsymbol{L}$) published so
far~\citep{sp34,sp6,RITnew,faunew}, except two binaries in~\citet{RITnewest}, which, when reproduced
by us, seem affected by imprecisions (M.~Jasiulek 2009, private communication). In particular, we plot the angle between 
$\boldsymbol{S}_{\rm fin}$ and $\boldsymbol{J}(r_{\rm in})$ as a
function of the spin-spin coupling ``magnitude''
$|\boldsymbol{a}_1 \times \boldsymbol{a}_2|$ for the binaries with $|\boldsymbol{a}_1 \times \boldsymbol{a}_2|\neq0$ (lower panel), and of $q$
for the binaries with $|\boldsymbol{a}_1 \times \boldsymbol{a}_2|=0$ (upper panel). Clearly,
assumption \textit{(iii)} is valid to within $4-5$ degrees,
with maximum errors of $7-8$ degrees. These errors are compatible with
the accuracy with which the NR simulations can compute the final-spin
direction for strongly precessing binaries (\textit{cf.} the error's noisy dependence on $|\boldsymbol{a}_1 \times \boldsymbol{a}_2|$).

\smallskip
\noindent\textit{(v) When the initial spin vectors are equal and
  opposite and the masses are equal, the spin of the final BH is the
  same as for nonspinning binaries}.
 Besides being physically reasonable -- reflecting the
expectation that if the spins are equal and opposite, their
contributions cancel out -- this assumption is confirmed by 
NR simulations [\cf discussion in~\citet{lr09}] 
and by the leading-order PN spin-spin and spin-orbit couplings. 

\smallskip

With these assumptions, we derive an expression for the final
spin. Let us write~\eqref{ell_def}, using \textit{(i)}, as
\begin{equation}
\label{assumption_1_bis}
\boldsymbol{a}_{\rm fin}=\frac{1}{(1+q)^2}
\left(\boldsymbol{a}_1(r)+\boldsymbol{a}_2(r)q^2 + 
\boldsymbol{{\ell}}(r) q \right)\,,
\end{equation}
where $\boldsymbol{a}_{\rm fin}= \boldsymbol{S}_{\rm fin}/M^2$ and
$\boldsymbol{{\ell}} \equiv \boldsymbol{\tilde{\ell}}/(M_1 M_2)$.
Using \textit{(ii)}, the final-spin norm is 
\begin{eqnarray}
\label{eq:general}
&\vert \boldsymbol{a}_{\rm fin}\vert=
\frac{1}{(1+q)^2}\Big[ \vta{1}^2 + \vta{2}^2 q^4+
 2 {\vert \boldsymbol{a}_2\vert}{\vert 
\boldsymbol{a}_1\vert} q^2 \cos \alpha\,+
\nonumber\\ 
& \hskip 0.2cm
2\left(
     {\vert \boldsymbol{a}_1\vert}\cos \beta(r) +
     {\vert \boldsymbol{a}_2\vert} q^2  \cos \gamma(r)
\right) {\vert \boldsymbol{{\ell}} \vert}{q}+\vert \boldsymbol{{\ell}}\vert^2 q^2
\Big]^{1/2},
\end{eqnarray}
where $\alpha, \beta$ and $\gamma$ are 
\begin{equation}
\label{cosines}
\cos \alpha \equiv
{\boldsymbol{\hat{a}}_1\cdot\boldsymbol{\hat{a}}_2}
\,,
\hskip 0.3cm
\cos \beta \equiv
 \boldsymbol{\hat a}_1\cdot\boldsymbol{\hat{{\ell}}}\,,
\hskip 0.3cm
\cos \gamma \equiv
\boldsymbol{\hat{a}}_2\cdot\boldsymbol{\hat{{\ell}}}\,.
\end{equation}
Note that because of $\textit{(iv)}$ $\alpha$ does not depend on the
separation and is simply the angle between the spins at the \textit{initial} separation $r_{\rm in}$.
The angles $\beta$ and $\gamma$ are instead functions of the binary's separation, 
but this dependence cancels out in the linear combination in which they appear
in~\eqref{eq:general}, which is indeed, within the assumptions made, independent of
the separation and which can therefore be evaluated at $r=r_{\rm in}$. To see this, let us consider
expression~\eqref{eq:general} at the effective ISCO. There,
$\boldsymbol{J}_{_{\rm rad}}(r_{_{\rm ISCO}})=0$ by definition and
therefore $\boldsymbol{\ell}(r_{_{\rm ISCO}})=\boldsymbol{L}(r_{_{\rm
    ISCO}})$. As a result, $\beta(r_{_{\rm ISCO}})$ [$\gamma(r_{_{\rm
      ISCO}})$] are simply the angles between
$\boldsymbol{S}_1$ [$\boldsymbol{S}_2$] and $\boldsymbol{L}$ at the ISCO. Using
now assumption $\textit{(iv)}$, we can write part
of~\eqref{eq:general} as
\begin{align}
\label{assumption3eq}
&\vta{1}\cos\beta(r_{_{\rm ISCO}}) + \vta{2} q^2 \cos\gamma(r_{_{\rm
     ISCO}}) = (\boldsymbol{\hat L}\cdot \boldsymbol{S})_{_{\rm
     ISCO}}/M_1^2 \nonumber\\
&\hskip 0.5cm =(\boldsymbol{\hat L}\cdot
 \boldsymbol{S})/M_1^2=\vta{1}\cos\widetilde{\beta}(r) + \vta{2} q^2
 \cos\widetilde{\gamma}(r)\,,
\end{align}
where $\widetilde{\beta}$ and $\widetilde{\gamma}$ are the angles
between the spins and $\boldsymbol{L}$ at \textit{any separation} $r$
and thus also at $r=r_{\rm in}$
\begin{equation}
\label{cosines2}
\cos \widetilde{\beta} \equiv
 \boldsymbol{\hat a}_1\cdot\boldsymbol{\hat{{L}}}\,,
\hskip 0.5cm
\cos \widetilde{\gamma} \equiv
\boldsymbol{\hat{a}}_2\cdot\boldsymbol{\hat{{L}}}\,.
\end{equation}

This proves our previous statement: the dependence on $r$
that $\beta$ and $\gamma$ have in
expression~\eqref{eq:general} is canceled by the linear combination in
which they appear. Stated differently, the final-spin norm is
simply given by expression~\eqref{eq:general} where $\beta(r)
\rightarrow \widetilde{\beta}(r_{\rm in})$ and $\gamma(r) \rightarrow
\widetilde{\gamma}(r_{\rm in})$. Thus, one does not need to
worry about the angles between $\boldsymbol{\hat a}_{1,2}$ and
$\boldsymbol{\hat{{\ell}}}$ but simply about the angles between $\boldsymbol{\hat a}_{1,2}$ and
$\boldsymbol{\hat{{L}}}$ at $r=r_{\rm in}$, which are easy to compute.

Finally, we need to compute $\vert \boldsymbol{{\ell}}\vert$
and for this we proceed like in \citet{old} and match
expression~\eqref{eq:general} at $r=r_{_{\rm ISCO}}$
with~\eqref{eqspin_uneqmass} for parallel and aligned spins
[$\alpha=\beta(r_{_{\rm ISCO}})=\gamma(r_{_{\rm ISCO}})=0$], for
parallel and antialigned spins [$\alpha=0$, $\beta(r_{_{\rm
      ISCO}})=\gamma(r_{_{\rm ISCO}})=\pi$], and for antiparallel
spins which are aligned or antialigned [$\alpha=\beta(r_{_{\rm
      ISCO}})=\pi$, $\gamma(r_{_{\rm ISCO}})=0$ or
  $\alpha=\gamma(r_{_{\rm ISCO}})=\pi$, $\beta(r_{_{\rm
      ISCO}})=0$]. As noted in \citet{old}, this matching is
not unique, but the degeneracy can be broken by exploiting assumption
\textit{(v)} (\ie by imposing that $\vtl$ does not depend on
$\boldsymbol{a}_{1,2}$ when $\boldsymbol{a}_{1} = -
\boldsymbol{a}_{2}$ and $q=1$) and by requiring for simplicity that
$\vtl$ depends linearly on $\cos\alpha$, $\cos\beta$ and
$\cos\gamma$. Using these constraints and~\eqref{assumption3eq} we
obtain again an expression valid for \textit{any separation} and hence
for $r=r_{\rm in}$
\begin{eqnarray}
\label{eq:L2}
&&\vtl = 2 \sqrt{3}+ t_2 \nu + t_3 \nu^2 +\\
&&
 \frac{s_4}{(1+q^2)^2} \left(\vta{1}^2 + \vta{2}^2 q^4 
	+ 2 \vta{1} \vta{2} q^2 \cos\alpha)\right) + 
\nonumber \\
&&
\left(\frac{s_5 \nu + t_0 + 2}{1+q^2}\right)
	\left(\vta{1}\cos\widetilde{\beta}(r_{\rm in}) + 
        \vta{2} q^2 \cos\widetilde{\gamma}(r_{\rm in})\right)\,.
\nonumber 
\end{eqnarray}

When comparing~\eqref{eq:general},~\eqref{assumption3eq}
and~\eqref{eq:L2} with (8) and (11) of~\citet{old}, it is
straightforward to realize that they are mathematically the same,
although derived under a different (and improved) set of
assumptions. This 
is simply because the new assumptions \textit{(ii)} and \textit{(iv)} are
compatible, via equation~\eqref{assumption3eq}, with the old
ones. However, the new assumptions make a substantial difference in
the prediction of the final-spin \textit{direction}. Since
\textit{(iii)} states that $\boldsymbol{a}_{\rm fin} \parallel
\boldsymbol{J}(r_{\rm in})$, the angle $\theta_{\rm fin}$ between the
final spin and the initial orbital angular momentum
$\boldsymbol{L}(r_{\rm in})$ is given by
\begin{equation}
\label{eq:angle} \cos \theta_{\rm fin} =
\boldsymbol{\hat{L}}(r_{\rm in}) \cdot \boldsymbol{\hat{J}}(r_{\rm
  in})\,.  
\end{equation} 
This expression replaces and improves~(10) of~\citet{old} and, as we
will show, is verified both for initial separations of a few
gravitational radii, such as those considered in NR, and for larger
separations (\eg $\sim 10^4\,M$), which are relevant for cosmological
applications. 

\begin{figure*}
\centerline{
\resizebox{8.cm}{!}{\includegraphics[angle=0]{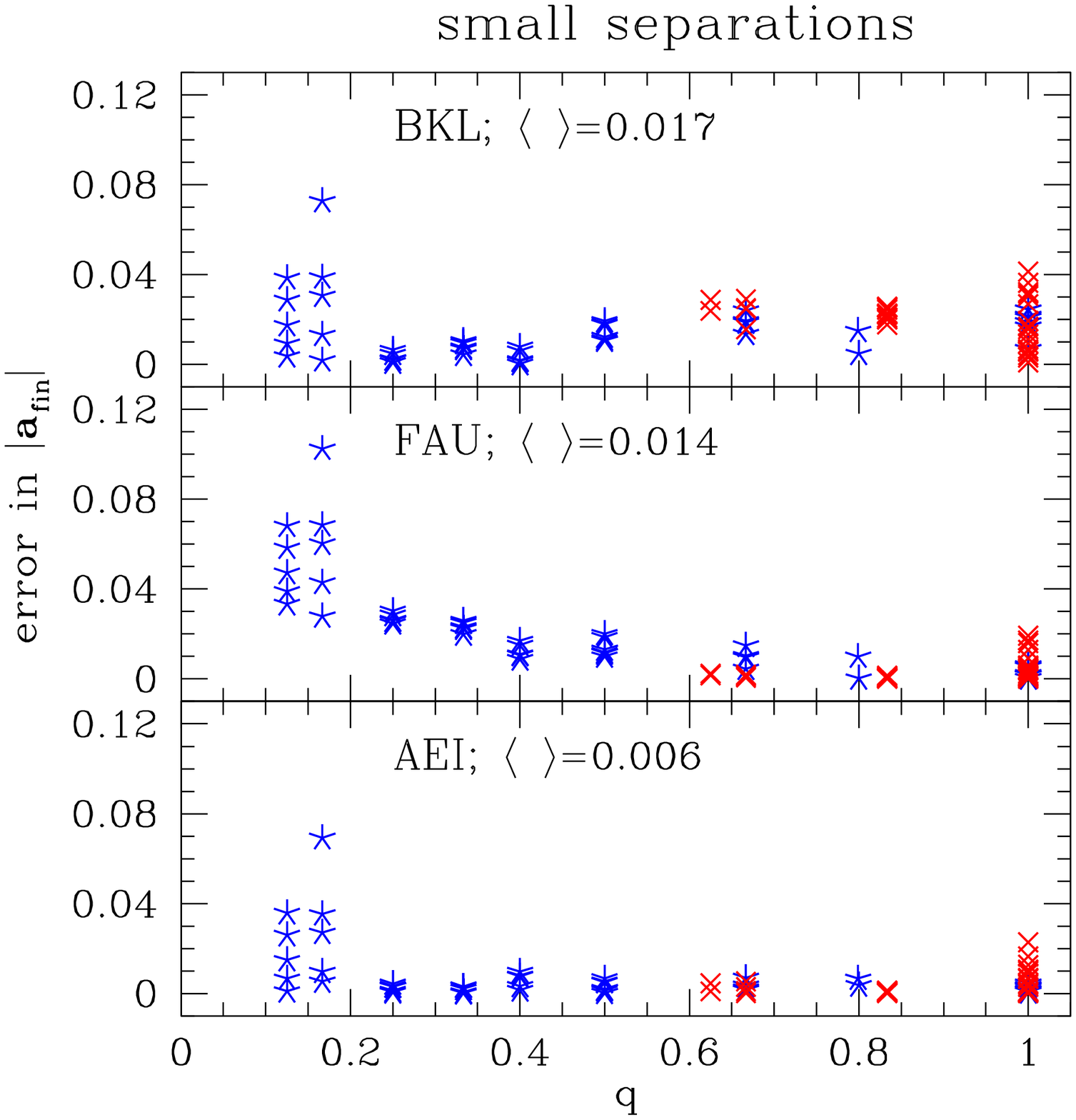}}
\hskip 1.0cm
\resizebox{8.cm}{!}{\includegraphics[angle=0]{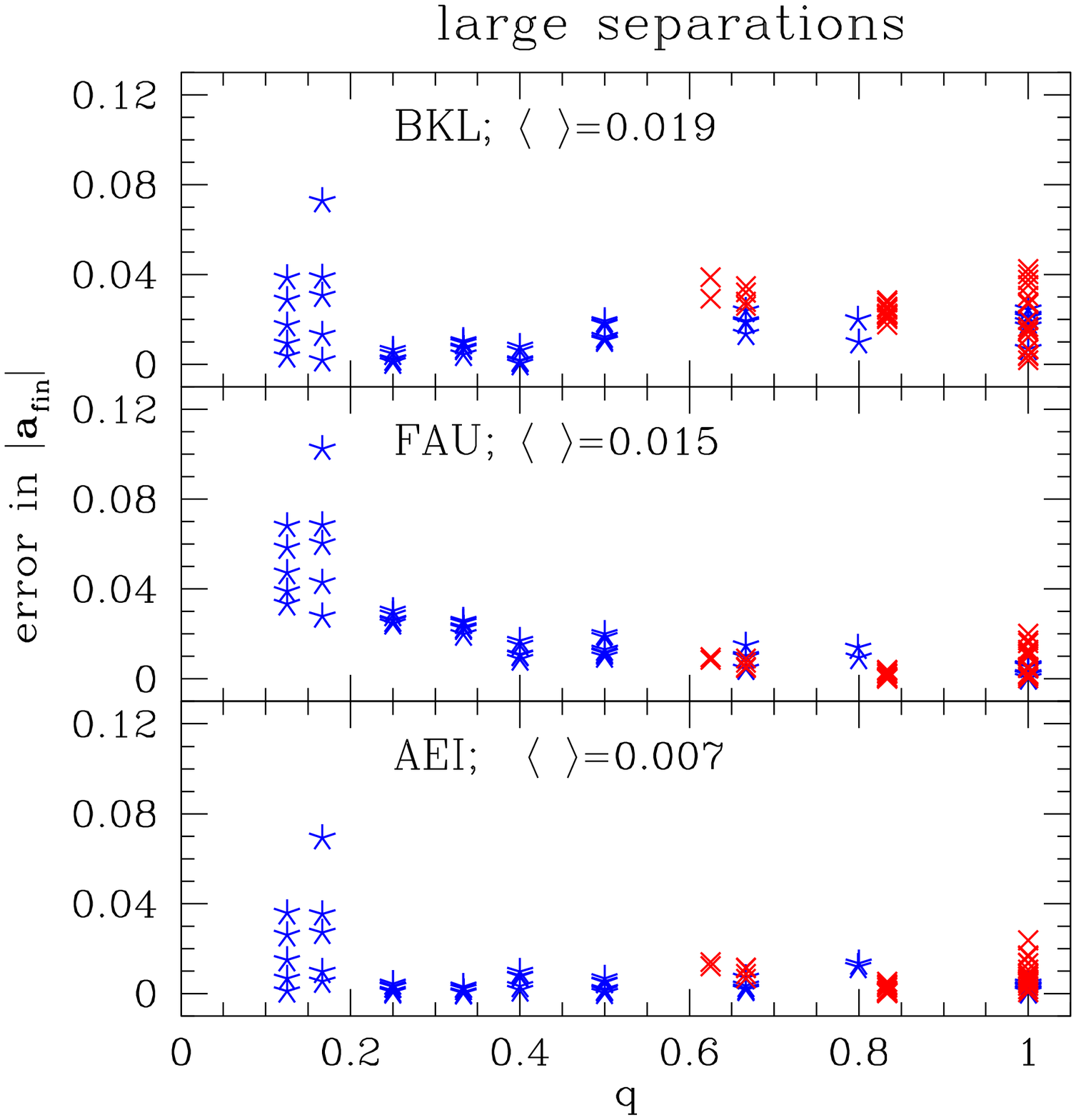}}
}
\caption{\textit{Left panel:} error in the final-spin norm
 $\vert\vert\boldsymbol{a}_{\rm fin,NR}\vert-\vert\boldsymbol{a}_{\rm fin, *}\vert\vert$ (``*'' being
  either ``AEI'', ``FAU'' or ``BKL'') as a function of the mass ratio, when the various formulas are applied to the
  \textit{small-separation} configurations corresponding to the
  initial data of the 45 simulations
  of~\citet{RITnew},~\citet{sp34,sp6} (blue stars), and to the 41 simulations of~\citet{faunew,fau} (red crosses). The numbers next to each
 label indicate the average error. \textit{Right panel:} the same as in
  the left one but for binaries at \textit{large separations} of $r= 2\times 10^4 M$. Although the new AEI expression is slightly better, all
  formulas give accurate predictions for
   $\vert\boldsymbol{a}_{\rm fin}\vert$ both for small and large separations.
  The larger errors for small $q$ are most likely due to the larger
  truncation errors affecting those simulations~\citep{RITnew}.}
\label{fig:afin}
\end{figure*}

\begin{figure*}
\centerline{
\resizebox{8.cm}{!}{\includegraphics[angle=0]{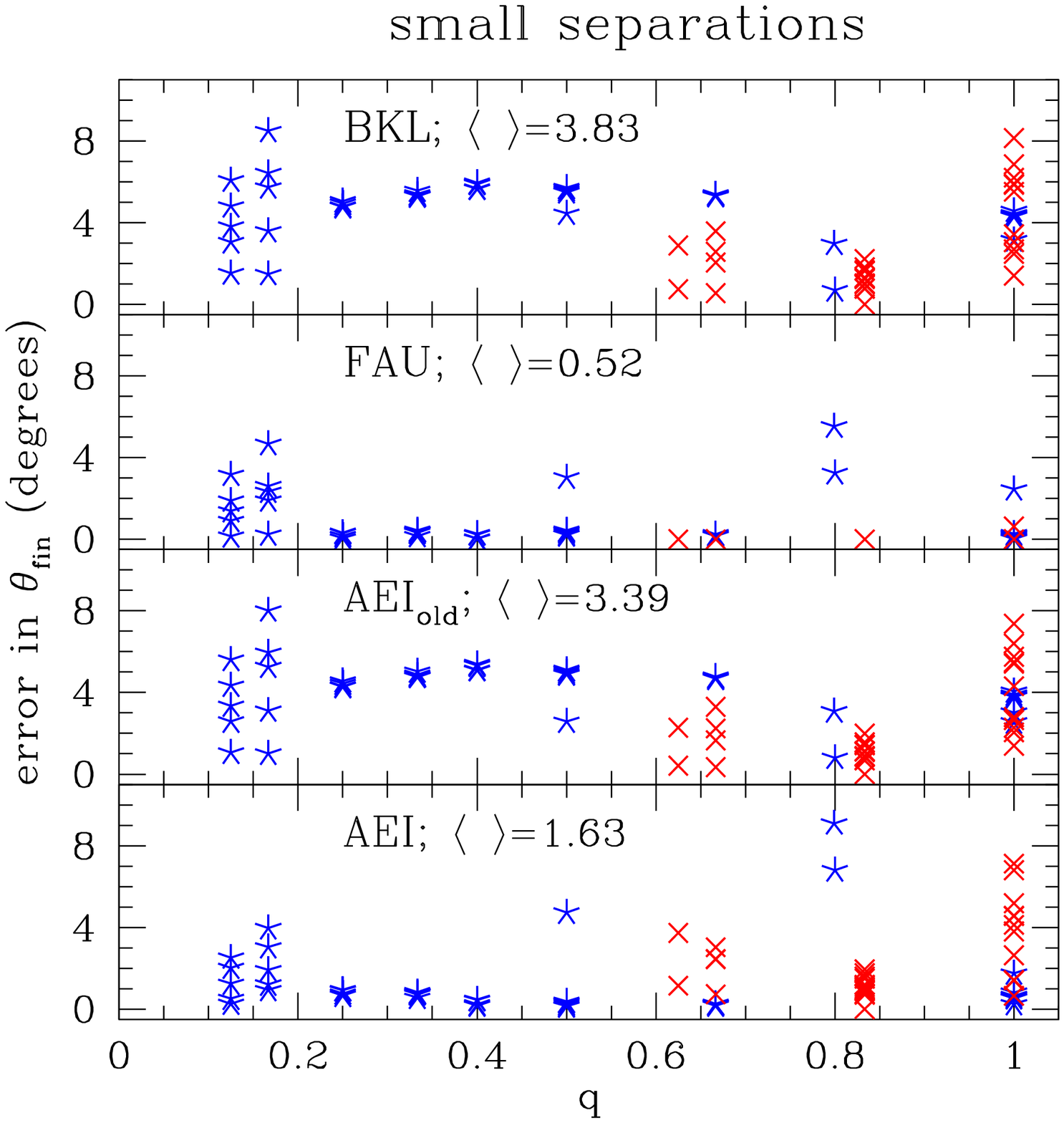}}
\hskip 1.0cm
\resizebox{8.cm}{!}{\includegraphics[angle=0]{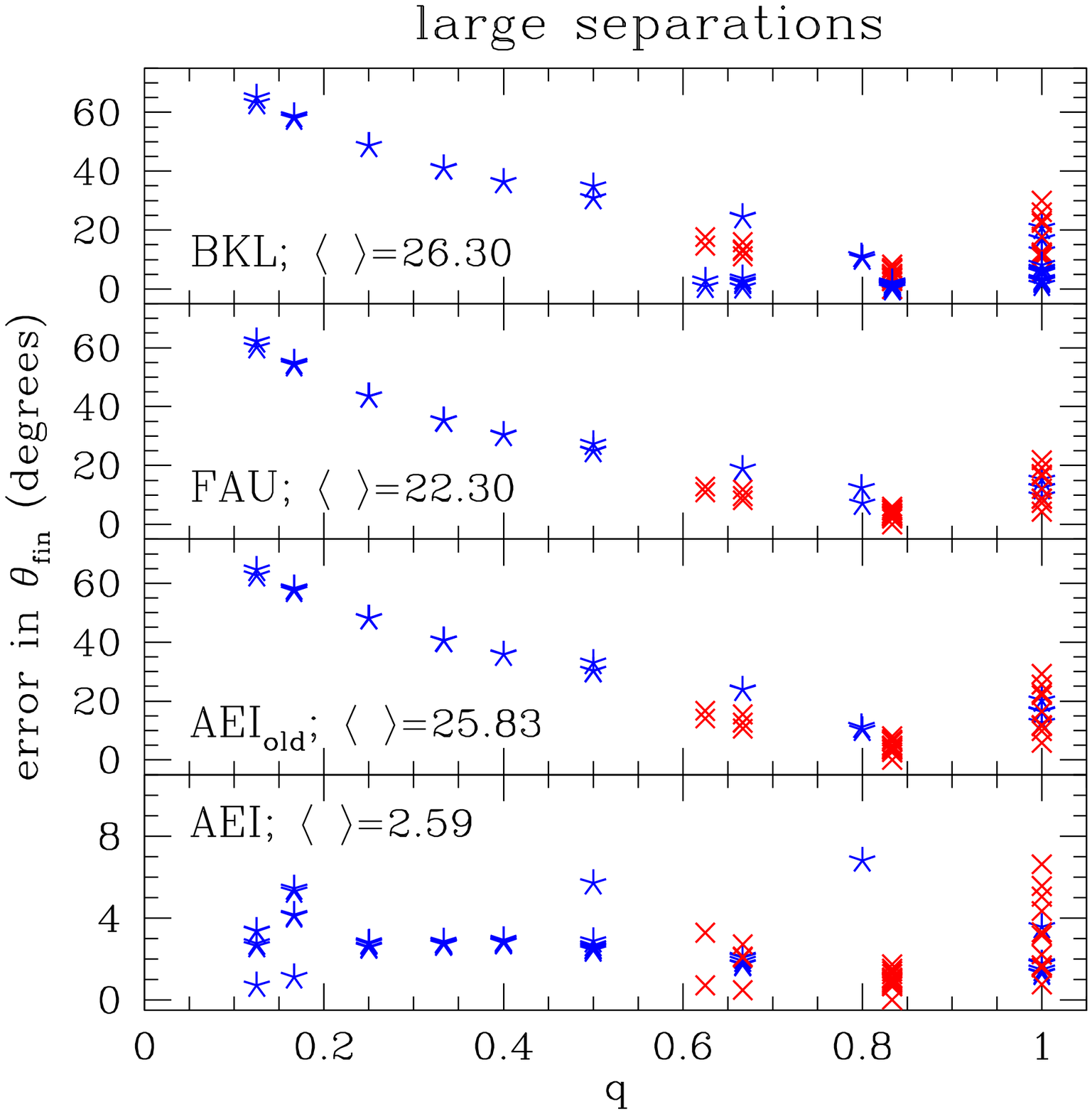}}
}
\caption{
Same as in Fig.~\ref{fig:afin}, but
 for the inclination-angle error
 $\vert\theta_{\rm fin, *}-\arccos{[\boldsymbol{\hat{L}}(r_{\rm in}) \cdot \boldsymbol{\hat{a}}_{\rm
fin,NR}]}\vert$ and without the 8 binaries of~\citet{fau}, for which the
 final-spin direction was not published. Here too, the larger errors for small-separation binaries with small $q$ are likely due to the larger
  truncation errors. The new AEI
  expression is accurate both for small and large separations, while
  the other ones become imprecise for large separations.}
\label{fig:thetafin}
\end{figure*}

\section*{Comparison with numerical relativity data}

We now test our expressions~\eqref{eq:general},~\eqref{assumption3eq} and~\eqref{eq:L2}
for $|\boldsymbol{a}_{\rm fin}|$ and our expression~\eqref{eq:angle} for $\theta_{\rm fin}$ 
against the NR simulations published so far for \textit{generic}
binaries. Note that to test the expression for $\theta_{\rm fin}$ we
use the same simulations as in Fig.~\ref{fig:test_assumption3} [\ie \citet{sp34,sp6,RITnew,faunew}], while to test the expression for $|\boldsymbol{a}_{\rm fin}|$ we also use the data of~\citet{fau} (which only reports the final-spin norm).
Also, we compare our 
predictions (AEI) with those of similar formulas suggested
by~\citet{bkl} (BKL),~\citet{old} (${\rm AEI_{old}}$)
and~\citet{faunew} (FAU). The comparison consists of two
steps. First, we use as
input the initial data of the NR simulations, in which the binaries have 
\textit{small separations} (\ie $r_{\rm in}\lesssim 10\, M$). 
Second, we use binaries at \textit{large separations} (\ie $r_{\rm in} \leq
2 \times 10^4\, M$), for which the dynamics starts being dominated by gravitational-wave emission and which are therefore
of direct relevance for cosmological investigations.
More precisely, we evolve the NR initial configurations back in time up to a separation of $2 \times 10^4\, M$ using the 2.5 PN
equations in the quasi-circular limit~\citep{buonanno_vallisneri}, calculating the predictions of
the different formulas at each step and considering the maximum error for each formula. 
(We stress, however, that the figures we present do not change significantly 
if we integrate only up to a separation of $\sim 200 M$.)

The left panel of Fig.~\ref{fig:afin} shows the predictions of the various formulas
for $\vert\boldsymbol{a}_{\rm fin}\vert$, when applied to the small-separation
configurations corresponding to the initial data of the NR simulations. 
In particular, it reports the
error in the final-spin norm $\vert\vert\boldsymbol{a}_{\rm fin,
  NR}\vert-\vert\boldsymbol{a}_{\rm fin, *}\vert\vert$ -- where ``*''
stands for ``AEI'' (which, as already stressed, 
gives the same predictions for $\vert\boldsymbol{a}_{\rm fin}\vert$ 
as ``${\rm AEI_{old}}$''), ``FAU'' or ``BKL'' -- as a function of $q$.
The right panel shows instead the \textit{maximum}
error when the configurations are evolved back in time up to $r_{\rm in}= 2\times 10^4 M$. 
Fig.~\ref{fig:afin} clearly shows
that although the AEI expression is slightly better, one \textit{can}
apply the old formulas to widely-separated binaries \textit{if} the
focus is on  $|\boldsymbol{a}_{\rm fin}|$ only. Note that the larger errors
for small $q$ are most likely due to the larger truncation errors
affecting those simulations [see~\citet{RITnew}, sec. IIIA].

However, the situation is very different for the final-spin
\textit{direction}. In particular, 
Fig.~\ref{fig:thetafin} reports the inclination-angle error $\vert\theta_{\rm fin, *}-\arccos{[\boldsymbol{\hat{L}}(r_{\rm in}) \cdot \boldsymbol{\hat{a}}_{\rm
  fin,NR}]}\vert$,
for all the data in Fig.~\ref{fig:afin}, except those of~\citet{fau}, 
for which the final-spin direction was not
published. When considering small-separation binaries (left panel), our new formula
performs slightly better than the ``BKL'' and ``${\rm AEI_{old}}$'' formulas, but it is not better than the ``FAU'' one. Indeed,
the latter is exact by construction for the 32 binaries of Table II in~\citet{faunew}, because for such data the final-spin direction
was not published and has been here reconstructed
using the FAU formula. When applied to 
large-separation binaries (right panel), however, our new formula performs much better
than the other ones. 
This result is not surprising since all the previous formulas
predict the same value of $\theta_{\rm fin}$ for all $r_{\rm in}$, 
as they neglect the precession of the binary's orbital plane.
As suggested by~\cite{fau}, one can in principle use PN theory to integrate the binary to small separations, and then
apply the old formulas. As already mentioned, however, this is impractical, 
compared to our simple and algebraic expressions for the final spin.
Also, note that the error in $\theta_{\rm fin}$ made by the
previous formulas at large separations increases for small-$q$ binaries. This is again
because the previous formulas predict the same $\theta_{\rm fin}$ for all $r_{\rm in}$. 
Therefore, because the correct $\theta_{\rm fin}$ becomes small for large $r_{\rm in}$ 
[since $\cos(\theta_{\rm fin, NR}) \approx \boldsymbol{\hat{L}}(r_{\rm in}) \cdot \boldsymbol{\hat{J}}(r_{\rm in})\approx 1$], 
the maximum error of the previous formulas is roughly given by their prediction for 
 $\theta_{\rm fin}$ at small separations, which can be large for 
small $q$ if the angle between $\boldsymbol{\hat{S}}$ and $\boldsymbol{\hat{L}}$ at small separations is large [because for $q\approx0$, 
$ \boldsymbol{\hat{J}}(r_{\rm in})\approx\boldsymbol{\hat{S}}(r_{\rm in})$ 
and $\cos(\theta_{\rm fin}) \approx \boldsymbol{\hat{L}}(r_{\rm in}) \cdot \boldsymbol{\hat{S}}(r_{\rm in})$].

The very small errors in the predictions of the final-spin direction also
provide additional evidence, besides that in Fig.~\ref{fig:test_assumption3}, of
the  validity of assumption \textit{(iii)}, namely that 
${\boldsymbol{\hat{S}}}_{\rm fin} \simeq {\boldsymbol{\hat{J}}}(r_{\rm
  in})$. Also, they suggest a correlation
between the final-spin direction and the orbital plane when the
binary was widely separated. Stated differently, by observing $\boldsymbol{\hat{a}}_{\rm fin}$,
\eg via a jet if this is assumed along $\boldsymbol{\hat{a}}_{\rm fin}$, one is virtually ``observing'' $\boldsymbol{\hat{J}}(r_{\rm in})$
and can conclude that the orbital plane at large separations was roughly orthogonal
to the final spin. Our result could therefore be applied to
X-shaped radio sources, for which the origin of the double pair of
jets is under debate~\citep{capetti,jets2}.

\section*{Conclusions}

We have derived a new formula predicting the spin of
the BH resulting from the merger of two BHs in quasi-circular orbits
and having arbitrary initial masses and spins. Our derivation is based
on a revised set of assumptions and exploits an additional constraint
to reduce to only \textit{four} the number of undetermined
coefficients. The new formula is identical to that proposed
in~\citet{old} in the prediction of the final-spin \textit{norm}, but is different in the
prediction of its \textit{direction},
showing a much better agreement with the numerical data. The new formula can be applied to binaries with
separations larger than $\sim200 M$ \textit{without}
any preliminary integration of the PN equations, in contrast with what
would be needed by the other formulas proposed in the literature. Thus, our formula is particularly suitable for
astrophysical and cosmological applications and could provide clues about the relation between the spin of the SMBH in the
center of AGNs and the binary's  orbital plane well before the
merger.

\acknowledgements
We are grateful to M. Jasiulek, L. A. Gergely and A. Buonanno for helpful
discussions. E.B. acknowledges support from NSF grant PHY-0603762.

\end{document}